\documentclass[%
superscriptaddress,
 preprint,
showkeys,showpacs,preprintnumbers,
amsmath,amssymb,
aps,
prb,
]{revtex4-2}

\usepackage{graphicx}
\usepackage{dcolumn}
\usepackage{bm}

\usepackage{upgreek}
\usepackage{amsmath}

\begin{document}


\title{Metasurface-based all-optical diffractive convolutional neural networks}

\author{Zhijiang Liang}
\affiliation{School of Information Engineering, Nanchang University, Nanchang 330031, China}

\author{Chenxuan Xiang}
\affiliation{School of Information Engineering, Nanchang University, Nanchang 330031, China}

\author{Shuyuan Xiao}
\email{syxiao@ncu.edu.cn}
\affiliation{School of Information Engineering, Nanchang University, Nanchang 330031, China}
\affiliation{Institute for Advanced Study, Nanchang University, Nanchang 330031, China}

\author{Jumin Qiu}
\affiliation{School of Physics and Materials Science, Nanchang University, Nanchang 330031, China}

\author{Jie Li}
\affiliation{College of Optoelectronic Engineering, Chengdu University of Information Technology, Chengdu 610225, China}

\author{Qiegen Liu}
\affiliation{School of Information Engineering, Nanchang University, Nanchang 330031, China}

\author{Chengjun Zou}
\email{zouchengjun@ime.ac.cn}
\affiliation{Institute of Microelectronics, Chinese Academy of Sciences, Beijing 100029, China}

\author{Tingting Liu}
\email{ttliu@ncu.edu.cn}
\affiliation{School of Information Engineering, Nanchang University, Nanchang 330031, China}
\affiliation{Institute for Advanced Study, Nanchang University, Nanchang 330031, China}

\begin{abstract}

The escalating energy demands and parallel-processing bottlenecks of electronic neural networks underscore the need for alternative computing paradigms. Optical neural networks, capitalizing on the inherent parallelism and speed of light propagation, present a compelling solution. Nevertheless, physically realizing convolutional neural network (CNN) components all-optically remains a significant challenge. To this end, we propose a metasurface-based all-optical diffractive convolutional neural network (MAODCNN) for computer vision tasks. This architecture synergistically integrates metasurface-based optical convolutional layers, which perform parallel convolution on the optical field, with cascaded diffractive neural networks acting as all-optical decoders. This co-design facilitates layer-wise feature extraction and optimization directly within the optical domain. Numerical simulations confirm that the fusion of convolutional and diffractive layers markedly enhances classification accuracy, a performance that scales with the number of diffractive layers. The MAODCNN framework establishes a viable foundation for practical all-optical CNNs, paving the way for high-efficiency, low-power optical computing in advanced pattern recognition.

\end{abstract}

\maketitle


\section{\label{sec1}Introduction}

Artificial intelligence (AI), which seeks to simulate and extend human intelligence, has been profoundly transformed by artificial neural networks (ANNs). By emulating the interconnected structure of biological neurons, ANNs excel at representation learning and pattern recognition, finding widespread application in domains such as image analysis, healthcare, and industrial automation\cite{LeCun2015}. However, as these models scale in depth and complexity, electronic implementations face unsustainable energy consumption and fundamental bottlenecks in parallel computing efficiency. In this context, optical neural networks (ONNs) have emerged as a promising alternative, leveraging the inherent parallelism and speed of light to overcome these limitations\cite{Wetzstein2020,Shastri2021,Zhou2022}.

ONNs utilize light as the information carrier, constructing computational architectures with a variety of optical components. Signal processing is achieved through the modulation, propagation, and interference of optical fields between neuron layers. This paradigm offers intrinsic advantages, including light-speed parallel processing, low power consumption, and the absence of electro-optical conversion latency, positioning it as a key solution to the compute-energy dilemma of electronic systems\cite{Shen2017,Cheng2024}. Among various implementations, diffractive neural networks (DNNs) have shown remarkable potential since this pioneering concept was introduced for object classification\cite{Lin2018,Yan2019} and subsequently for advanced AI tasks\cite{Li2023,Zhan2024,Xiong2024,Bai2024,Qiu2024,Wang2025,Zhou2025,Chen2025,Qiu2025}. Despite their promise, conventional DNNs fabricated via 3D printing face challenges in device miniaturization and operation at shorter wavelengths (e.g., visible and near-infrared), limiting their practical deployment. Moreover, diffractive elements generally offer less versatile optical field control compared to emerging flat-optics platforms, hindering multi-dimensional multiplexing and high-density integration.

Metasurfaces, composed of subwavelength nanostructures, provide unprecedented control over the phase\cite{Deng2018,Li2024}, amplitude\cite{Zheng2021,Deng2024,Liu2024}, polarization\cite{Arbabi2019,Ding2020,Wang2023}, and orbital angular momentum\cite{Ren2020, Meng2025} of light, making them a more efficient platform for implementing advanced optical AI than conventional diffractive optical elements. Notable demonstrations include non-interleaved dielectric metasurfaces for multi-channel polarization imaging\cite{Deng2020,Li2020,Liu2022,Ou2022,Yuan2023} and on-chip metasurface-based DNNs for parallel multi-task computing in the visible range\cite{Luo2022,Luo2023,He2024,Xiang2025}. To enhance feature extraction specificity—a strength of convolutional neural networks (CNNs)—integrating optical convolution kernels with DNNs has become a critical research direction. Early hybrid systems, such as those incorporating diffractive optical elements as convolutional layers, demonstrated the potential for low-power processing\cite{LeCun2010,Guo2025,Chang2018}, while recent meta-optics-based convolutional accelerators have paved the way for miniaturized, high-speed hardware\cite{Zheng2022,Luo2024,Zheng2024,Fu2022}. Nevertheless, a fully integrated, all-optical architecture that seamlessly combines metasurface-based convolution with metasurface-based diffraction remains largely unexplored.

To bridge this gap, we propose a metasurface-based all-optical diffractive convolutional neural network (MAODCNN). This architecture integrates metasurface-configured convolutional kernels for direct, efficient feature extraction (e.g., edges and textures) from coherent optical fields, overcoming the limitations of diffraction-only networks. The convolved features are then fed directly into a metasurface-based diffractive network for inference, establishing a complete photoelectric-conversion-free pipeline. We validate MAODCNN on image classification tasks using the MNIST and Fashion-MNIST datasets. Results show an 12\% improvement in classification accuracy over pure diffractive networks, attributable to the precise feature capture by the all-optical convolutional kernels. Looking forward, we envision the optoelectronic co-packaging of metasurface convolutional layers, diffractive layers, and detector arrays into millimeter-scale chips. Such integrated modules could be deployed in miniaturized endoscopes or multifunctional edge-computing sensors, providing critical supports for the practical realization of all-optical intelligent systems.

\section{\label{sec2}Working Principle}

The proposed MAODCNN is schematically illustrated in Fig. 1a. The architecture comprises an input layer (for encoding images into coherent optical fields), a metasurface-based convolutional layer, multiple hidden layers formed by diffractive metasurfaces, and an output layer consisting of a photodetector array. Its operation mirrors the training paradigm of electronic neural networks but is executed entirely in the optical domain. Input images are first encoded as coherent optical fields. The metasurface convolutional layer then performs all-optical convolution for preliminary feature extraction. Subsequently, the diffractive layers, optimized via an angular spectrum propagation algorithm and error backpropagation, iteratively process the optical field through interference and diffraction. This process directs the optical energy from different input classes to distinct, pre-assigned regions on the output detector plane. The final classification is determined by comparing the mean optical intensity received by each detector region.

\begin{figure*}[htbp]
	\centering
	\includegraphics[width=\linewidth]{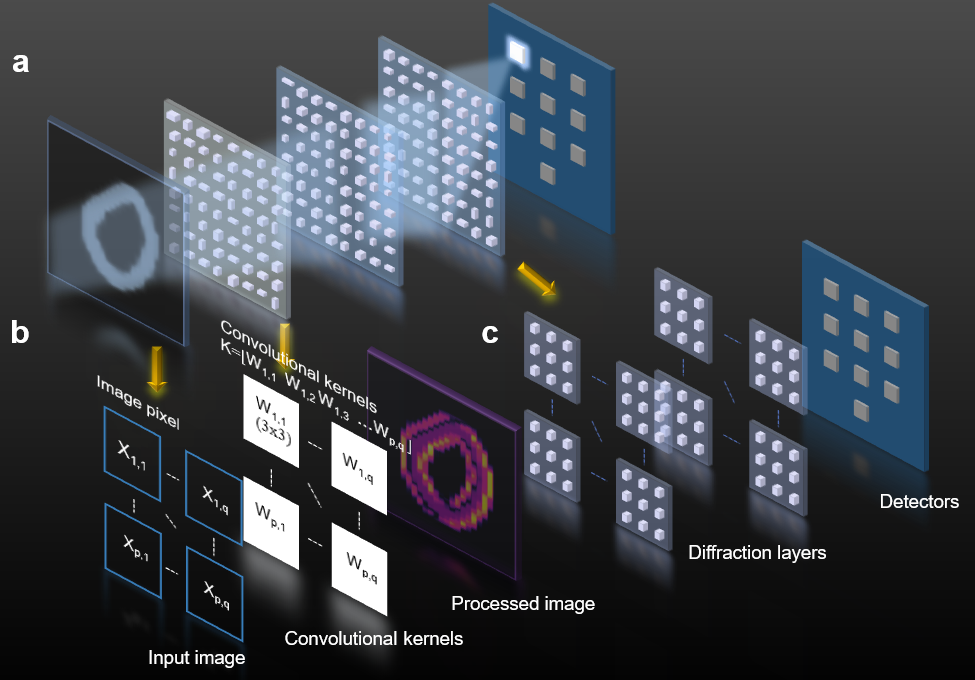}
	\caption{Architecture of the metasurface-based all-optical diffractive convolutional neural network (MAODCNN). (a) Schematic of the full process of all-optical classification, from optical input to the final decision at the detector array. (b) Design of the metasurface convolutional layer, which performs parallel convolution by tiling a single kernel design (comprising $n$ sub-segments) across the aperture. (c) Configuration of the cascaded diffractive metasurface layers and the output detector plane.}
	\label{fig1}
\end{figure*}

CNNs excel at hierarchical feature extraction through convolutional layers, pooling layers, and fully connected layers\cite{Yamashita2018,Tian2018}. The convolutional layer, in particular, employs learnable kernels that slide across the input, leveraging weight sharing to reduce parameters while preserving spatial correlations. In our work, we implement these convolutional kernels using metasurfaces. As shown in Fig. 1b, our designed metasurface, with dimensions of $p\times q$ segments, directly processes a $p\times q$-pixel image. To extract multiple features in parallel, each segment is subdivided into $n$ sub-segments, each functioning as a $3\times3$ convolutional kernel. Every kernel comprises $3\times3$ meta-units that modulate the phase or amplitude of the incident light, with their values optimized via deep learning for specific tasks. Upon illumination, the input optical field is modulated by these kernel weights. The results from all nine units within a kernel are instantaneously summed via optical interference, outputting a feature map rich in low-level information (e.g., edges and textures). This light-speed, parallel operation fundamentally overcomes the sequential bottleneck of electronic computing.

The diffractive layer is constructed from cascaded, transmissive metasurfaces, as depicted in Fig. 1c. The subwavelength elements on each metasurface act as artificial neurons, with learnable complex-valued transmission coefficients. Following the Huygens-Fresnel principle, each neuron connects to all neurons on the subsequent layer via diffraction, functioning as a secondary wave source. The modulation imposed by each neuron is analogous to the function of a weight in an electronic network. During training, we employ an error backpropagation algorithm (stochastic gradient descent) with datasets like MNIST and Fashion-MNIST to iteratively optimize the phase parameters of all metasurface layers. The training objective is to ensure that, after multi-layer diffraction, the optical energy from a given input class is focused onto its corresponding designated area on the detector array. Physically, the phase modulation is encoded in the geometry of the metasurface's subwavelength structures. No active components are required between layers, enabling passive, low-power all-optical classification.

As a proof of concept, we adopt a metasurface unit cell based on TiO$_{2}$ nanocolumns on a SiO$_{2}$ substrate (Fig. 2a). This design offers a high refractive index contrast and compatibility with standard nanofabrication processes. By meticulously tuning the nanocolumn dimensions ($D_{x}$, $D_{y}$) and height ($H$) at a fixed period ($P=400$ nm) for a wavelength of 532 nm, we can achieve efficient phase and amplitude control. We perform systematic simulations using the finite-difference time-domain (FDTD) method. The transmittance and phase maps in Fig. 2b and 2c, respectively, demonstrate that by sweeping $D_{x}$ and $D_{y}$, we can identify specific unit cell geometries that simultaneously provide target transmittance levels and a full 0 to $2\pi$ phase coverage. This establishes a critical foundation for accurately mapping convolutional kernel weights to physical metasurface structures, thereby verifying the feasibility of our design for implementing all-optical convolution and constructing the MAODCNN.

\begin{figure}[htbp]
	\centering
	\includegraphics[width=\linewidth]{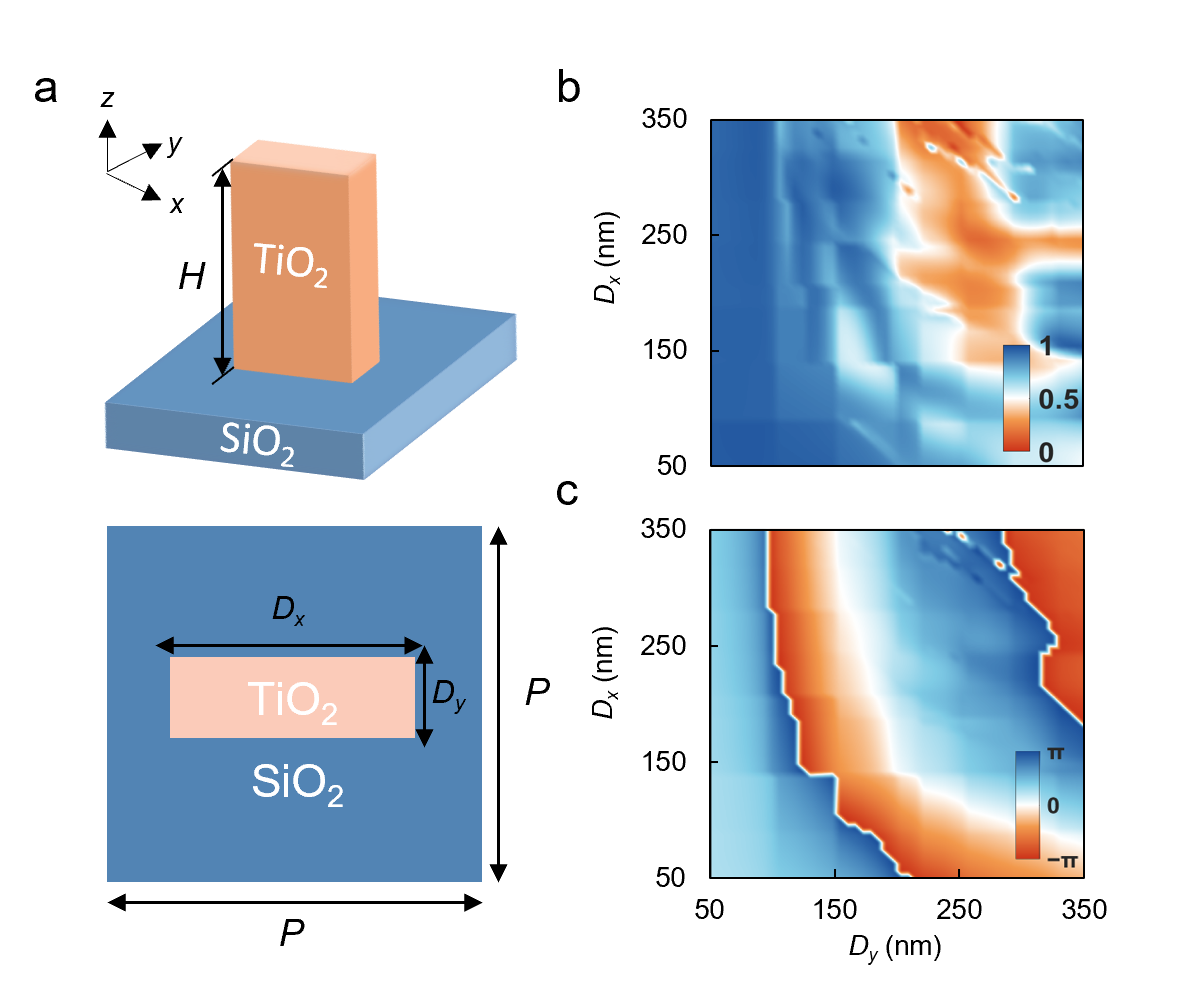}
	\caption{Metasurface unit cell design and optical characterization. (a) Schematic of a TiO$_{2}$ metasurface unit cell, showing the fixed period ($P$) and height ($H$), and tunable lateral dimensions ($D_{x}$, $D_{y}$) that define its optical response. (b), (c) FDTD-simulated transmission amplitude (b) and phase shift (c) as functions of $D_{x}$ and $D_{y}$ at a wavelength of 532 nm ($P=400$ nm, $H=600$ nm).}
	\label{fig2}
\end{figure}

\section{\label{sec3}Results and discussion}

To evaluate the performance of MAODCNN, we conduct numerical experiments on the MNIST and Fashion-MNIST datasets. The output plane is divided into 10 discrete regions, each corresponding to a data category. The category of an input image is determined by identifying the region with the highest optical intensity. We first investigate the impact of network depth using the MNIST test set (10,000 images). As shown in Fig. 3a, classification accuracy improves significantly with an increasing number of diffractive hidden layers (each with $280\times280$ neurons and a 400 nm array period). For the ten-class task, the accuracy of a two-hidden-layer configuration is 20\% higher than that of a single-layer network. This demonstrates a clear ``depth advantage" for MAODCNN, even in the absence of optical nonlinearity. Furthermore, the results indicate that for classification tasks with fewer categories, only one or two hidden layers are sufficient to meet the requirements, highlighting the model's architectural efficiency. A comparative analysis with the DNN on the MNIST dataset (Fig. 3b) reveals the superior performance of MAODCNN. With a single hidden layer, MAODCNN achieves a 15\% higher classification accuracy than DNN. This underscores the advantage of incorporating a dedicated convolutional layer for more robust and hierarchical feature extraction, leading to enhanced adaptability to data noise and deformation.

\begin{figure}[htbp]
	\centering
	\includegraphics[width=\linewidth]{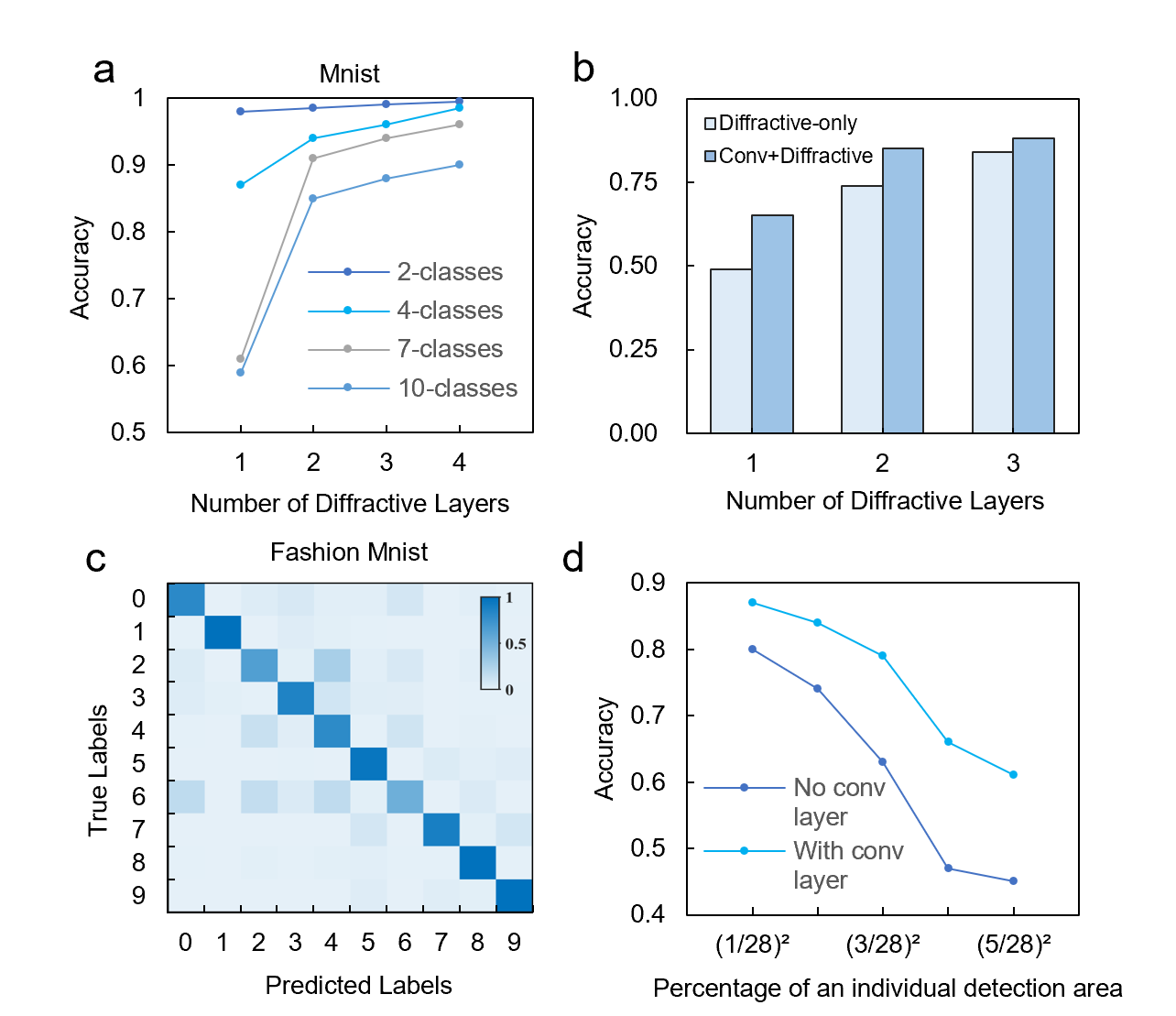}
	\caption{Performance evaluation of MAODCNN on MNIST and Fashion-MNIST datasets. (a) Classification accuracy on MNIST as a function of the number of hidden layers and output classes. (b) Comparative classification accuracy between a conventional DNN and the proposed MAODCNN on the 10-class MNIST task. (c) Confusion matrix of the MAODCNN's predictions on the Fashion-MNIST test set. (d) Classification accuracy on MNIST versus the occupancy rate of a single detection region for DNN and MAODCNN.}
	\label{fig3}
\end{figure}

The model's capability on more complex datasets is validated by the confusion matrix for Fashion-MNIST (Fig. 3c). The matrix shows favorable classification accuracy despite the dataset's higher feature complexity and inter-class similarity, intuitively visualizing misclassifications and confirming the model's efficacy in challenging scenarios. We further analyze the influence of detector design by varying the occupancy rate of a single detection region when processing $280\times280$ inputs (Fig. 3d). The results indicate that a smaller detection region significantly enhances recognition accuracy. For a direct performance comparison between MAODCNN and DNN, a detector specification with a single-region occupancy rate of 2/28 is selected for subsequent experiments.

In contrast to conventional diffractive networks that rely solely on phase modulation, MAODCNN incorporates additional amplitude modulation, raising the need to assess its potential impact. We evaluate both a purely phase-modulated network and a network incorporating amplitude crosstalk on the MNIST test set. The output light field distributions of these two networks are compared in Figs. 4a and 4b, visually illustrating their similar modulation effects. Quantitatively, the recognition error between the two is negligible (Fig. 4c). Furthermore, an analysis of the normalized energy distribution across all digit classes (``0" to ``9") shows that the average energy in the target regions exceeds 25\%, with minimal deviation (error bars around 3\%) from the phase-only network (Fig. 4d). These results confirm that the impact of amplitude crosstalk on the system's classification performance is negligible, as phase modulation dominates the optical field manipulation.

\begin{figure}[htbp]
	\centering
	\includegraphics[width=\linewidth]{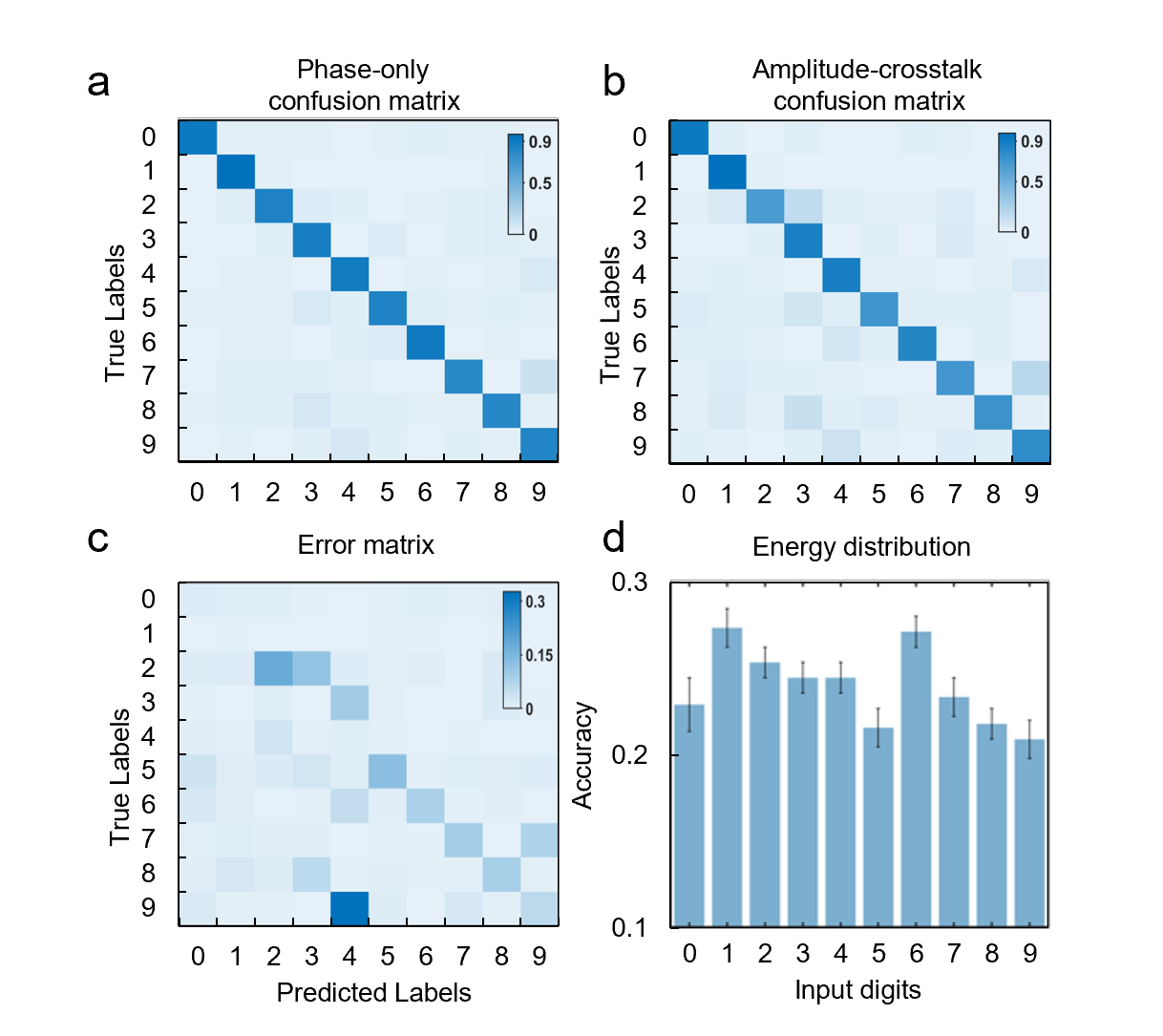}
	\caption{Impact of amplitude crosstalk on the MAODCNN performance. (a) Output light field distribution of the purely phase-modulated MAODCNN. (b) Output light field distribution of the MAODCNN incorporating amplitude crosstalk. (c) Recognition error between the phase-only network and the network with amplitude crosstalk across the test set. (d) Normalized energy distribution for each digit class (``0"-``9") in the MAODCNN with amplitude crosstalk, demonstrating minimal deviation (error bars ~3\%) from the phase-only network.}
	\label{fig4}
\end{figure}

A direct architectural comparison between DNN and MAODCNN is summarized in Table 1. While DNN processes the $280\times280$ input directly through two diffractive layers, MAODCNN first transforms the input into feature maps via an optical convolutional layer of the same scale. This key architectural difference underpins the performance gain of MAODCNN and provides a basis for optimizing future optical neural networks. 

\begin{table*}[htbp]
	\centering  
	\caption{Architectural comparison between DNN and MAODCNN models.}  
	\label{tab:architecture}  
	\includegraphics[width=\linewidth]{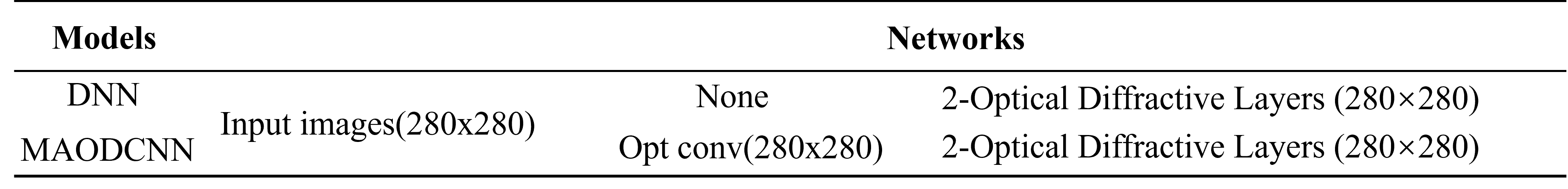}  
\end{table*}

Representative class prediction scores and activation maps for both datasets are shown in Fig. 5. For the MNIST dataset (e.g., digit ``3"), the model accurately redistributes the input energy to the detection region as expected, with key activation regions (white squares) precisely localizing the digit features. For Fashion-MNIST (e.g., shoes), the model also shows a clear advantage for the correct category, effectively capturing essential characteristics like shape and contour, thus demonstrating robust adaptability across diverse visual tasks. The functionality of the metasurface convolutional kernel is exemplified in Fig. 6 using the handwritten digit ``0". The output image exhibits a ring-like structure with markedly enhanced edge contrast and sharpness compared to the input. This ``edge sharpening" effect is achieved as the phase-only kernel modulates the complex amplitude of the input by adjusting the cosine (real) and sine (imaginary) components of the phase. This specific modulation in the complex domain enhances intensity contrast at edges through optical interference and superposition, thereby validating the kernel's effectiveness for primary feature enhancement.

\begin{figure}[htbp]
	\centering
	\includegraphics[width=\linewidth]{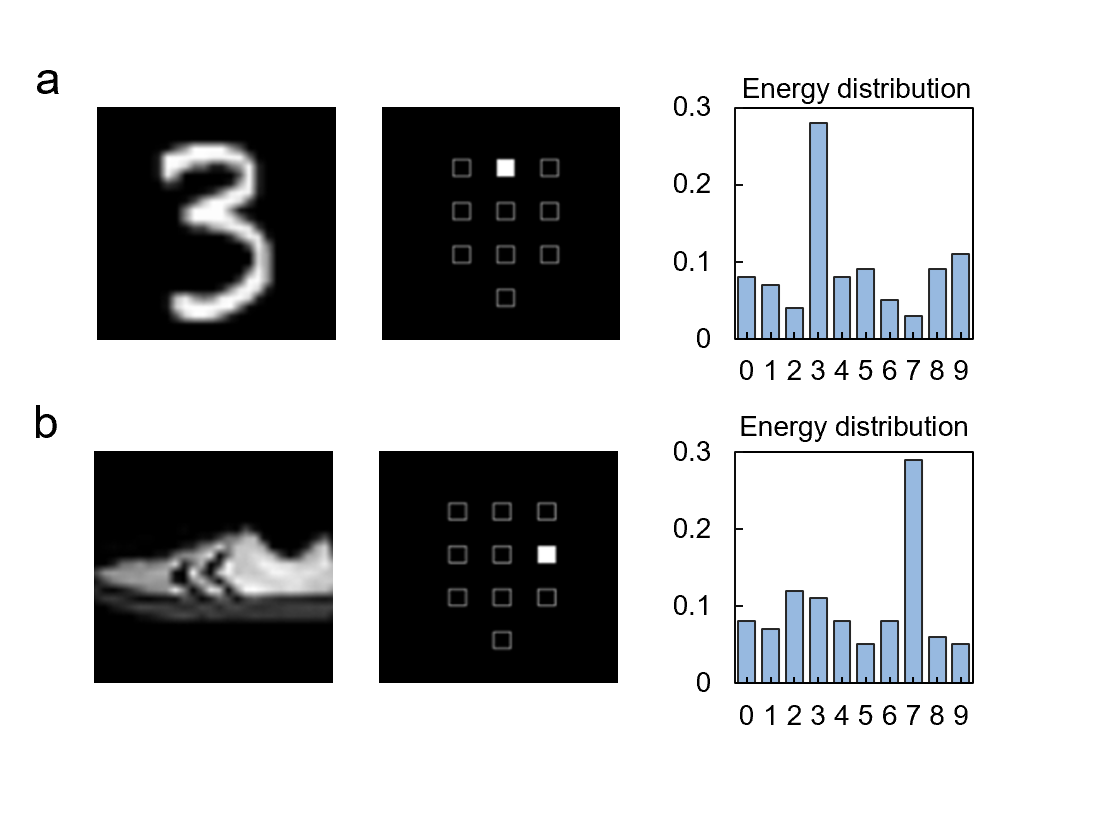}
	\caption{Representative output patterns and class activation. Simulated output optical patterns and the corresponding normalized energy distribution across the 10 detector regions for (a) an input handwritten digit ``3" (MNIST) and (b) an input ``shoe" (Fashion-MNIST). The white squares highlight the activated detector region corresponding to the correct class.}
	\label{fig5}
\end{figure}

\begin{figure}[htbp]
	\centering
	\includegraphics[width=\linewidth]{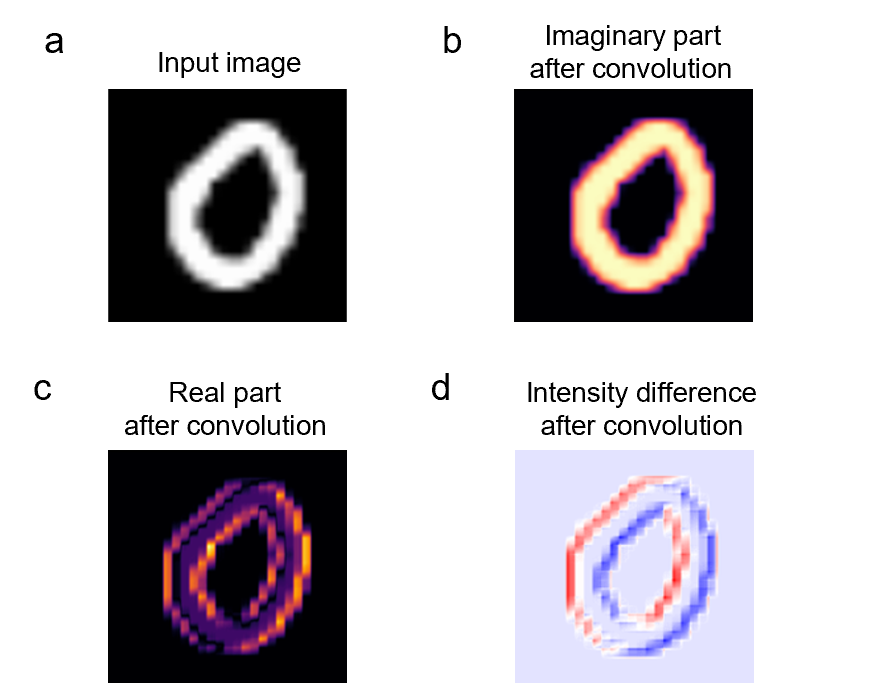}
	\caption{Feature map demonstrating edge enhancement. (a) The original handwritten digit ``0" before convolution. (b)The real component of the output light field after the input passes through the phase-only metasurface convolutional kernel. (c)The imaginary component of the output light field after convolution, showing its unique pattern. (d) The intensity difference distribution of the output light field after convolution.}
	\label{fig6}
\end{figure}

The training accuracy curves on both MNIST and Fashion-MNIST datasets further substantiate the architectural advantage of MAODCNN (Fig. 7). On the MNIST dataset, MAODCNN's accuracy not only converges more rapidly but also stabilizes at a significantly higher level (86\%) compared to DNN (74\%). This performance trend is consistently replicated on the more complex Fashion-MNIST dataset, where MAODCNN achieves a final accuracy of 79\%, outperforming DNN (73\%). The faster convergence and superior final performance across both datasets demonstrate that the integration of convolutional operations facilitates more efficient feature extraction and pattern learning than what can be achieved by diffractive layers alone. This underscores the critical role of the convolutional module in enhancing the training efficacy and classification performance of diffractive neural networks.

\begin{figure}[htbp]
	\centering
	\includegraphics[width=\linewidth]{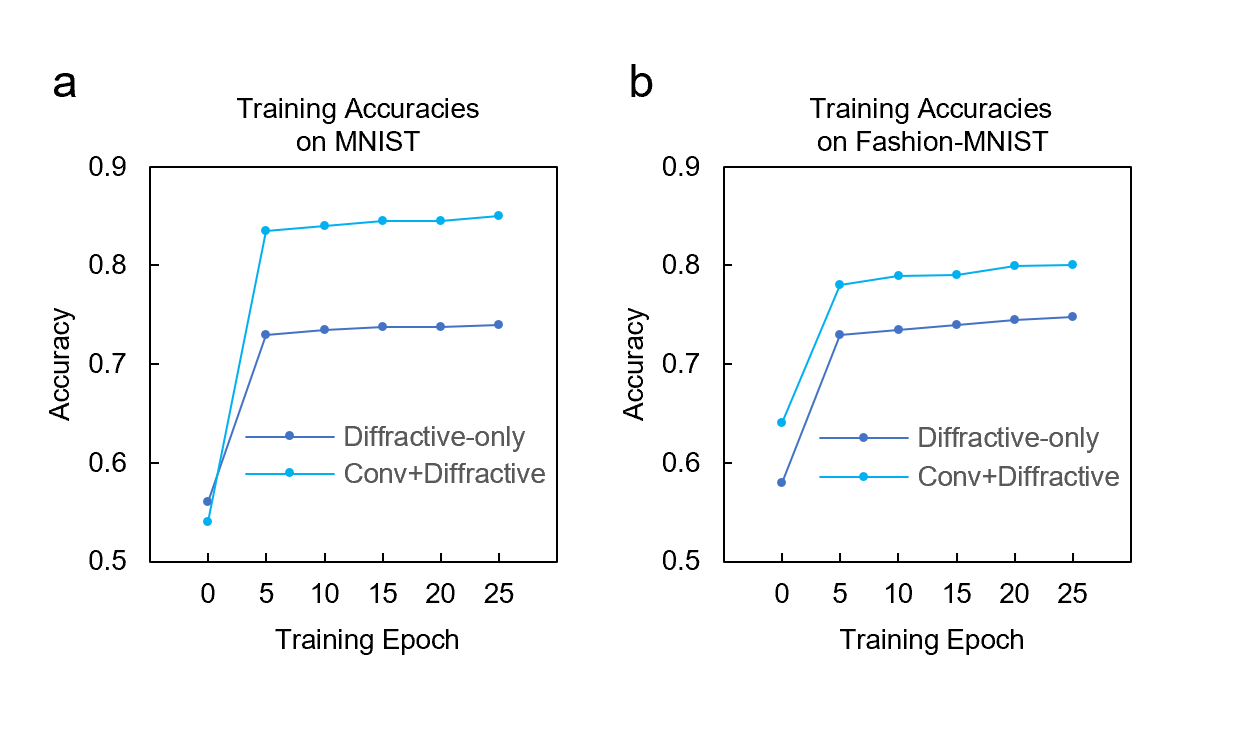}
	\caption{Training dynamics and performance comparison. Classification accuracy of MAODCNN versus DNN throughout the training process on the (a) MNIST and (b) Fashion-MNIST datasets.}
	\label{fig7}
\end{figure}

\section{\label{sec4}Conclusions}

In conclusion, we have established a complete all-optical processing pipeline—MAODCNN—by synergistically integrating a phase-only metasurface convolutional layer with diffractive neural networks, thereby offering a viable solution to the latency and energy bottlenecks inherent in electronic computing. Numerical validation on MNIST and Fashion-MNIST classification tasks demonstrates that MAODCNN significantly outperforms conventional DNNs, achieving an accuracy improvement of up to 15\% with only a single hidden layer. This result unequivocally confirms the critical role of dedicated optical convolution in enhancing feature extraction and classification performance. Looking forward, while the phase-only design proves effective, exploring joint phase-amplitude modulation in metasurfaces presents a promising avenue to further enrich feature representation for more complex visual tasks. The proposed MAODCNN architecture provides a foundational and practical framework for developing next-generation, high-efficiency optical intelligent chips, charting a clear path toward overcoming the limitations of electronic neural networks.

\begin{acknowledgments}	
	
This work was supported by the National Natural Science Foundation of China (Grants No. 12364045, No. 62305372, No. 12404484, No. 12264028, and No. 12304420), the Natural Science Foundation of Jiangxi Province (Grants No. 20232BAB201040 and No. 20232BAB211025), the Sichuan Science and Technology Program (Grants No. 2025NSFSC2073 and No. 2025ZNSFSC0846), and the Young Elite Scientists Sponsorship Program by JXAST (Grants No. 2023QT11 and No. 2025QT04). 

Zhijiang Liang and Chenxuan Xiang contributed equally to this work.
		
\end{acknowledgments}


%

\end{document}